\documentclass[lettersize,journal]{IEEEtran}
\usepackage{amsmath,amsfonts}
\usepackage{algorithmic}
\usepackage[ruled,lined]{algorithm2e}
\usepackage{array}
\usepackage[caption=false,font=normalsize,labelfont=sf,textfont=sf]{subfig}
\usepackage{textcomp}
\usepackage{stfloats}
\usepackage{url}
\usepackage{balance}
\usepackage{tabularx}
\usepackage{multicol}
\usepackage{verbatim}
\usepackage{graphicx}
\usepackage{threeparttable}
\usepackage[colorlinks]{hyperref}
\hypersetup{citecolor=blue}
\usepackage{cite}
\begin{document}

%\title{Iterative Gaussian Approximation Decoder Design for Random Spreading Unsourced Random Access in Gaussian MAC}
%modified by Chae
\title{Iterative Gaussian Approximation for Random Spreading Unsourced Random Access}

% or this one?
%\title{Universal Iterative Gaussian Decoder for Random Spreading Unsourced Random Access in Gaussian MAC}

\author{Liandong Hu, Jian Dang,~\IEEEmembership{Senior Member,~IEEE}, Zaichen Zhang,~\IEEEmembership{Senior Member,~IEEE}
	\vspace{-5mm}
	
	\thanks{ }% <-this % stops a space
	\thanks{L. Hu, J. Dang, and Z. Zhang are with the National Mobile Communications Research Laboratory, Frontiers Science Center for Mobile Information Communication and Security, Southeast University, Nanjing, 210096, China. J. Dang, Z. Zhang are also with the Purple Mountain Laboratories, Nanjing 211111, China (e-mail: \{liandonghu, dangjian, zczhang\}@seu.edu.cn).}% <-this % stops a space
	\thanks{Corresponding authors: Jian Dang (dangjian@seu.edu.cn).}}



\maketitle
\begin{abstract}
%Massive machine-type communications (mMTC) require scalable and energy-efficient solutions for supporting dense connectivity. Unsourced random access (URA) has emerged as a promising paradigm to achieve these goals by avoiding user-specific coordination. Within URA, the random spreading (RS) approach offers strong anti-interference capabilities and near-capacity performance under low user activity. In this paper, we propose an iterative Gaussian approximation decoder designed for RS-URA systems. The decoder exchanges extrinsic and intrinsic soft information between a multi-user detector and the channel decoder, enabling rapid convergence within a few iterations. It is universally adaptable to different RS structures and seamlessly integrates with techniques such as power division. Simulation results demonstrate that the proposed decoder achieves robust decoding performance, outperforming existing RS-URA schemes (e.g., RS-Polar) by up to 4 dB gain while supporting large numbers of active users. Furthermore, the iterative structure maintains low complexity and scalability, making it suitable for practical deployment in mMTC scenarios.

Massive machine-type communications (mMTC) demand robust solutions to support extensive connectivity efficiently. Unsourced random access (URA) has emerged as a promising approach, delivering high spectral and energy efficiency. Among URA code structures, the random spreading (RS) category is a key enabler, providing strong anti-interference capabilities through spectrum spreading gain. Notably, RS-URA approaches theoretical performance limits over the Gaussian multiple access channel in scenarios with few active users. In this paper, we propose an iterative Gaussian approximation decoder designed universally for RS-URA categories. The proposed receiver iterates extrinsic and intrinsic soft information to enhance decoding performance, requiring only a few iterations to converge. Numerical results validate the decoder's effectiveness in terms of performance and robustness.

\end{abstract}


\section{Introduction}

\subsection{Background and Related Work}

Unsourced random access (URA) \cite{Polyanskiy} reformulates the finite block-length multiple access (MA) problem \cite{FBL-FAS1,FBL-FAS2} within a theoretical coding framework, enabling elegant achievability performance analysis~\cite{URA_zzt}. Unlike coordinated MA schemes, such as time-division multiple access (TDMA), ALOHA, treating interference as noise (TIN), etc., URA requires neither user scheduling nor identification, which leads to significant improvements in both energy and spectral efficiency~\cite{acdemia and industrial1}. These advantages are especially pronounced in high-activity scenarios, where coordinated MA either demands prohibitively high energy or fails to support massive connectivity due to excessive complexity. Recently, URA has manifested substantial superioerity in terms of intergated sensing and communication \cite{UNISAC0}, \cite{UNISAC1}, \cite{UNISAC2}, \cite{UNISAC3}, \cite{UNISAC4}.

The Gaussian multiple access channel (GMAC) is one of the most important application scenarios for URA~\cite{UMA_proc,Zhang2025VTC}, focusing on the design of basic MA concatenated codes. Its effectiveness has been validated under realistic channel conditions~\cite[Fig.~2b]{acdemia and industrial2}. Several promising GMAC URA schemes have been proposed, including sparse codes~\cite{acdemia and industrial2},~\cite{sparse1},~\cite{sparse2},~\cite{sparse3}, random spreading~\cite{RS-Polar},~\cite{RS-LDPC},~\cite{PD_MAC},~\cite{RS-DJ}, segmented schemes~\cite{RS-zzt},~\cite{CCS2, CRC-BMST}, among others. In particular, random spreading URA (RS-URA) shows strong anti-interference capabilities by leveraging the spectrum spreading structure.

% Each user in in GMAC URA framework projects a small fraction of its message sequence onto a codebook with random elements to generate a signature. Then, it encodes and modulates the remaining portion of the message sequence and spreads the modulated signal by the selected signature. 

 In the RS-URA structure, a small portion of each user's message sequence is treated as a header to select a signature from a common codebook, where each signature is randomly generated. The remaining bits are then processed through channel encoding, modulation, and spectrum spreading using the selected signature. RS-URA can also be feasibly extended to systems with other channel models~\cite{MIMO-SISO0},~\cite{MIMO-SISO1},~\cite{ MIMO-SISO2},~\cite{MIMO-SISO3}. %From a signal processing perspective, RS-URA enables the construction of multiple measurement vector (MMV) signals~\cite{RS-DJ}, which is particularly beneficial for key tasks such as activity detection.

\subsection{Challenges and Contributions}

Iterative channel decoding, a common technique in sparse codes \cite{UMA_proc}, iterates extrinsic and intrinsic soft information between multi-user detection and the channel decoder to enhance decoding performance, as exemplified in \cite[Eq.~5--Eq.~12]{Zhang2025VTC}. This approach unlocks diversities within the channel code domain, providing valuable feedback for overall decoding, such as interference cancellation in multiple access systems. However, its application to RS-URA has been underexplored.

A recent attempt to leverage channel code domain diversities is described in \cite[Section~IV-B]{RS-DJ}, where hard-verdict feedback information from the channel decoder updates the calculation of a posteriori constellation symbol probabilities, aiding multi-user detection in the corresponding algorithms. Nevertheless, this approach is constrained to a specific algorithm's structure and serves as an additional procedure to improve estimation precision, lacking universality.

In contrast, inspired by the turbo design in \cite{Zhang2025VTC}, this work derives essential components for an iterative decoder tailored for RS-URA, universally adaptable to all variant RS structures. The proposed decoder maintains flexibility across RS categories and seamlessly integrates techniques like power division. Moreover, the proposed receiver requires only few iteration to converge the output.

The content structure is as follows. The system model configurations are detailed in Section~\ref{sec.2}. The proposed iterative Gaussian approximation decoder is introduced in Section~\ref{sec.3}. Numerical results are presented in Section~\ref{sec.4}, and conclusions are provided in Section~\ref{sec.5}.

\section{System Model}
\label{sec.2}

We consider \( K_a \) single-antenna active users each aims to transmit $B$ bits of information to a single-antenna receiver over \( n \) real channel uses. 
The received signal, initially represented as a length-\(n_p n_c\) vector, can then be reordered into a matrix of size \(n_p \times n_c\) as 
\begin{align}\label{eq:1}
	\mathbf{Y} = \sum_{i\in \mathcal{D}}p_i\mathbf{a}_i\mathbf{x}_i+\mathbf{N},
\end{align}
	where $\mathcal{D}$ denotes the active indexes of the signature codebook with $|\mathcal{D}|=K_a$, $\mathbf{N}$ denotes additive white Gaussian noise (AWGN) with entries generated independently, each having zero mean and variance $\sigma_n^2$, $\mathbf{a}_i\in \mathbb{R}^{n_p}$ denotes the signature randomly selected by active users and $\mathbf{x}_i\in \mathbb{R}^{n_c}$ is the $i$-th user's signal to be spread. Meanwhile, to adopt the power division (PD) strategy, similar to~\cite{PD_MAC},~\cite{Zhang2025VTC},~\cite{Ahmadi_MSMRA}, we divide the columns of the signature codebook into \(m\) groups, assigning signatures in each group different power levels \(P_1, P_2, \dots, P_m\), with \((P_1 + P_2 + \dots + P_m)/m = 1\) and $p_i=\sqrt{P_i}$ is the signal amplitude that is assigned to the group to which $\mathbf{a}_i$ belong. The average per-user energy-per-bit is defined as
\begin{align}\label{eq:2}
  \frac{E_b}{N_0}&=\frac{n_c}{2B\sigma^2_n},
\end{align}
where signatures are assumed to have unit power and the spread constellations consumes total power of $n_c$. The performance metric is defined as the per-user probability of error (PUPE), defined as
\begin{equation}\label{eq:3}
	\text{PUPE}=\frac{1}{K_a} \sum_{\mathbf{x} \in \mathcal{L}} \mathbb{P}\left(\mathbf{x} \notin \tilde{\mathcal{L}}(\mathbf{Y})\right),
\end{equation}
where the set $\mathcal{L}$ contains messages of active users, $\tilde{\mathcal{L}}(\mathbf{Y})$ is the list of decoded message sequences.
\section{Description of Proposed Scheme}\label{sec.3}
\subsection{Encoder}

This section details the operational procedures of the RS encoder. Initially, each user splits its message sequence into two segments, containing \( B_p \) and \( B_c \) bits, respectively. The first segment is mapped to the columns of the signature codebook \( \mathbf{A} = \left[\mathbf{a}_1, \mathbf{a}_2, \ldots, \mathbf{a}_T\right] \in \mathbb{R}^{n_p \times T} \) to select a signature of length \( n_p \), where \( T = 2^{B_p} \). To construct \( \mathbf{A} \), its elements are first drawn independently from \( \mathcal{N}(0, 1) \), and then each column is normalized to unit norm, i.e., \( \|\mathbf{a}_i\|^2 = 1 \) for all \( i \in \{1, 2, \ldots, T\} \). 

Next, the second segment, comprising \( B_c \) information bits, is encoded using an appropriate forward error correction code and modulated via binary phase shift keying (BPSK) (\( 0 \rightarrow +1 \), \( 1 \rightarrow -1 \)) to produce the signal \( \mathbf{x}_i \in \{\pm1\}^{1 \times n_c} \). This signal is then spread using the user’s selected signature to generate the transmit signal, occupying \( n = n_c n_p \) channel uses.

\subsection{Iterative Decoder}
Following the spirit in \cite{Zhang2025VTC}, the decoding process is carried out iteratively in three main steps:
\begin{itemize}
	\item[-] First, the active signatures (i.e., the active columns of the signature codebook) are detected.
	\item[-] Next, an iterative decoder based on Gaussian approximation is applied.
	\item[-] Finally, SIC is performed to remove the contributions of the decoded messages from the received signal.
\end{itemize} 
Each of these steps is described in detail below.
\subsubsection{Active signature Detection}
\label{SECactivDec}
To perform iterative decoding, the receiver must first detect the active signatures. To achieve this, the received signal is modeled as a sparse linear regression problem:
\begin{equation}\label{eq:4prime}
	\mathbf{Y} = \mathbf{A}\mathbf{\Gamma}\mathbf{X}+\mathbf{N},
\end{equation}
where $\mathbf{\Gamma}$ is a diagonal matrix with only elements corresponding to the active signatures are ones and other diagonal elemtns are zeros, and $\mathbf{X}$ is a row-sparse matrix with rows corresponding to the active users are $\mathbf{x}_i$'s and other rows are zero. One advantage of using the RS structure is that the received signal model can be expressed in the form of \eqref{eq:4prime}, which corresponds to a multiple measurement vector (MMV) model. This enables the application of robust compressive sensing (CS) algorithms to obtain favorable solutions. These include correlation-based greedy matching pursuit methods \cite{RS-LDPC, RS-Polar, PD_MAC} or more advanced approximate message passing (AMP) algorithms \cite{CRC-BMST, CCS2, RS-DJ}, which are designed based on the signal's statistical features. For a performance comparison between the correlation-based and AMP-based algorithms, refer to \cite[Fig.~1]{RS-DJ}. 
\subsubsection{Iterative Decoder}\label{sec.3.2}
In the iterative decoder, soft log-likelihood ratio (LLR) information of the bits is exchanged iteratively between two main components: the elementary signal estimator (ESE) and the channel decoder (DEC). We use $l_{i,j}^{ese}$ and $l_{i,j}^{dec}$ to denote the LLR information of the $i$-th user's $j$-th encoded symbol corresponding to the ESE and DEC, respectively, where $i\in\mathcal{D},~j\in\{1,\ldots,n_c\}$. From $\eqref{eq:1}$, the received signal at the $j$-th channel use can be expressed as
\begin{equation}\label{eq:4}
	\begin{aligned}
		\mathbf{y}_j &= \sum_{i\in \mathcal{D}}p_i \mathbf{a}_ix_{i,j}+\mathbf{n}_j\\
		&= p_i \mathbf{a}_ix_{i,j}+\underbrace{\sum_{l\neq i}p_l \mathbf{a}_lx_{l,j}+\mathbf{n}_j}_{\mathbf{g}_{i,j}},
	\end{aligned}
\end{equation}
where $\mathbf{n}_j$ and $x_{i,j}$ are the $j$-th columns of $\mathbf{N}$ and $\mathbf{x}_i$, respectively, and $p_i \mathbf{a}_ix_{i,j}$ and $\mathbf{g}_{i,j}$ are the desired signal and interference-plus-noise terms. To obtain a sufficient statistic for detecting the desired scalar \(x_{i,j}\), we apply a {projection of the MMSE-filtered signal onto} \(\mathbf{a}_i\), i.e., we multiply both sides of \eqref{eq:4} by \(\mathbf{a}_i^T \mathbf{F}\), where \(\mathbf{F} = (\sigma_n^2 \mathbf{I}_{n_s} + \mathbf{A}_d \mathbf{A}_d^T)^{-1}\), and the columns of \(\mathbf{A}_d\) are the vectors \(p_i \mathbf{a}_i\) corresponding to the detected signatures. Thus, we obtain
\begin{equation}\label{eq:matched}
y_{i,j} = \mathbf{a}_i^T \mathbf{F} (p_i \mathbf{a}_i x_{i,j} + \mathbf{g}_{i,j}) = p_i \mathbf{a}_i^T \mathbf{F}\mathbf{a}_i x_{i,j} + \xi_{i,j},
\end{equation}
where $\xi_{i,j} = \mathbf{a}_i^T \mathbf{F}\mathbf{g}_{i,j}$.
By approximating the distribution of $\xi_{i,j}$ as Gaussian, the likelihood function can be written as
\begin{equation}\label{eq:5}
	p\left(y_{i,j}|x_{i,j}\right)\propto  \exp\left\{-\frac{(y_{i,j}  - p_i \mathbf{a}_i^T \mathbf{F}\mathbf{a}_i x_{i,j} - \mathrm{E}\left(\xi_{i,j}\right))^2}{\mathrm{Var}\left(\xi_{i,j}\right)}\right\}.
\end{equation}
Using the likelihood in \eqref{eq:5}, the ESE extrinsic LLRs can be derived as:
	\begin{align}
		\nonumber
		&l_{i,j}^{ese}=\ln \frac{p\left(y_{i,j}|x_{i,j}=+1\right)}{p\left(y_{i,j}|x_{i,j}=-1\right)}\\\nonumber
			&=\frac{(y_{i,j}  + p_i \mathbf{a}_i^T \mathbf{F}\mathbf{a}_i  - \mathrm{E}\left(\xi_{i,j}\right))^2}{\mathrm{Var}\left(\xi_{i,j}\right)}\\\nonumber
			&\quad\quad\quad\quad\quad\quad -\frac{(y_{i,j}  - p_i \mathbf{a}_i^T \mathbf{F}\mathbf{a}_i  - \mathrm{E}\left(\xi_{i,j}\right))^2}{\mathrm{Var}\left(\xi_{i,j}\right)}\\
&=\frac{4 p_i \mathbf{a}_i^T \mathbf{F}\mathbf{a}_i(y_{i,j}    - \mathrm{E}\left(\xi_{i,j}\right))}{\mathrm{Var}\left(\xi_{i,j}\right)}. \label{eq:6d}
	\end{align}
The iterative decoding procedure begins by initializing $\mathrm{E}(\xi_{i,j}) = 0$ and \textcolor{black}{$\mathrm{Var}(\xi_{i,j}) = \sum_{l \neq i} p_l^2 (\mathbf{a}_i^T \mathbf{F}\mathbf{a}_l)^2 + \sigma_n^2 \mathbf{a}_i^T\mathbf{F}\mathbf{F}^T\mathbf{a}_i$}, and proceeds with the following steps iteratively.
\begin{itemize}
	\item[] {\textbf{Step~1}: Generate $l_{i,j}^{ese}$ according to \eqref{eq:6d}}
	\begin{equation}\label{eq:7}
		\textcolor{black}{l_{i,j}^{ese}=\frac{4 p_i \mathbf{a}_i^T \mathbf{F}\mathbf{a}_i(y_{i,j}    - \mathrm{E}\left(\xi_{i,j}\right))}{\mathrm{Var}\left(\xi_{i,j}\right)}.}
	\end{equation}
	\item[] {\textbf{Step~2}: Feed $l_{i,j}^{ese}$ into the channel decoder and obtain the extrinsic LLR $l_{i,j}^{dec}$\footnote{The decoder should be a soft-output decoder that provides soft outputs.}
.}\\
	\item[] {\textbf{Step~3}: Subtract the intrinsic LLR $l_{i,j}^{ese}$ from the extrinsic LLR $l_{i,j}^{dec}$ to prevent error propagation from the previous iteration\footnote{For a performance comparison with and without LLR subtraction, please refer to \cite[Fig. 3]{IDMA}.}.}
	\begin{equation}\label{eq:8}
		l_{i,j}^{dec}\leftarrow l_{i,j}^{dec}-l_{i,j}^{ese}.
	\end{equation}
	\item[] {\textbf{Step~4}: Update $\mathrm{E}\left(x_{i,j}\right)$ and $\mathrm{Var}\left(x_{i,j}\right)$}
	\begin{subequations}\label{eq:9}
		\begin{align}
			\mathrm{E}\left(x_{i,j}\right)&=\tanh\left(\frac{l_{i,j}^{dec}}{2}\right),\label{eq:9a}\\
			\mathrm{Var}\left(x_{i,j}\right)&=1-\left(\mathrm{E}\left(x_{i,j}\right)\right)^{2}.\label{eq:9b}
			\end{align}
	\end{subequations}
	\item[]\textbf{Step~5}: Update $\mathrm{E}\left(\xi_{i,j}\right)$ and $\mathrm{Var}\left(\xi_{i,j}\right)$
	\begin{subequations}\label{eq:10}
		\begin{align}
			\mathrm{E}\left(\xi_{i,j}\right)&=\textcolor{black}{\sum_{l \in \mathcal{D} \backslash i}p_l\mathbf{a}_i^T\mathbf{F}\mathbf{a}_l\mathrm{E}\left(x_{l,j}\right)},\label{eq:10c}\\\nonumber
\mathrm{Var}\left(\xi_{i,j}\right)&=\sum_{l \in \mathcal{D} \backslash i}p_l^2\left(\mathbf{a}_i^T\mathbf{F}\mathbf{a}_l\right)^2\mathrm{Var}\left(x_{l,j}\right)\\
&\quad\quad\quad\quad+\sigma_n^2\mathbf{a}_i^T\mathbf{F}\mathbf{F}^T\mathbf{a}_i.\label{eq:10d}
		\end{align}
	\end{subequations}
\end{itemize}
{\em We should note that performance of the iterative decoder significantly depends on the channel code design in Step~2 of the algorithm.} However, while channel code design is a challenging research area, it is beyond the scope of this work.
\subsubsection{Successive Interference Cancellation}
\label{sec.3.3}
When the iterative decoder is stopped, SIC is applied by subtracting the signal components corresponding to the set of successfully decoded messages, denoted by $\tilde{\mathcal{L}}\left(\mathbf{Y}\right)$, which satisfy all bit-check conditions using appropriate verification methods. The SIC is performed as follows:
\begin{equation}\label{eq:11}
	\mathbf{Y}^{q} = \mathbf{Y}^{q-1} - \sum_{l\in \tilde{\mathcal{L}}\left(\mathbf{Y}\right)} p_l\mathbf{a}_l\mathbf{x}_l,
\end{equation}
where $ \mathbf{Y}^{q}$ denotes the residual received signal after applying SIC in the $q$th SIC round. Subsequently, $\mathbf{Y}^q$ is fed back to the active signature detection module for the next iteration. The general structure of the decoder is depicted in Algorithm~1.
\begin{algorithm}[!t]
{\small
    \caption{ Overview of the receiving algorithm.}
\label{algorithm1}
		\KwIn{Noisy observations $\mathbf{Y}$, codebook $\mathbf{A}$, upperbound for iterative decoding $\widehat{itr}$}
		\KwOut{Estimated message list $\tilde{\mathcal{L}}\left(\mathbf{Y}\right)$.}
\While{The set $\tilde{\mathcal{L}}\left(\mathbf{Y}\right)$ has changed from the previous iteration}
    {
\begin{enumerate}
    \item Perform activity detection according to Sec.~\ref{SECactivDec}.
\item Perform the iterative Gaussian approximation decoder\\ according to Sec.~\ref{sec.3.2}:
\begin{itemize}
    \item[]~Step~1: Generate $l_{i,j}^{ese}$ according to \eqref{eq:6d}.
    \item[]~Step~2: Obtain the extrinsic LLR $l_{i,j}^{dec}$ by passing \\
    $\quad\quad\quad \quad l_{i,j}^{ese}$ into the channel decoder.
    \item[]~Step~3: $l_{i,j}^{dec}\leftarrow l_{i,j}^{dec}-l_{i,j}^{ese}$.
    \item[]~Step~4: Update $\mathrm{E}\left(x_{i,j}\right) \& \mathrm{Var}\left(x_{i,j}\right)$ using~\eqref{eq:9}.
	\item[]~Step~5: Update $\mathrm{E}\left(\xi_{i,j}\right)\&\mathrm{Var}\left(\xi_{i,j}\right)$ using~\eqref{eq:10}.
    \end{itemize}
\item Perform SIC as according to Sec.~\ref{sec.3.3}:
\end{enumerate}
    }
    }
    \end{algorithm}
\section{Numerical Results}\label{sec.4}
	In this section, numerical results are presented to verify the effectiveness of the proposed URA scheme. The setups adopted in this paper are as follows: new radio low density parity check (NR-LDPC) with a code rate of $1/3$ is used as the channel code; the number of information bits is $B=100$; the pilot codebook size is $B_p=12$; the number of bits fed into the channel decoder is $B_c=88$; the total number of channel uses is $n=30,000$; the lengths of the channel codeword and pilot are $n_c=264$ and $n_p=114$, respectively. Simultaneous orthogonal matching pursuit (SOMP) \cite{SOMP0},~\cite{SOMP1},~\cite{SOMP} is used for activity detection due to its feasibility. The upper bound of iterative decoding is fixed to $\widehat{itr} = 20$. To exploit diversity in the power domain, signatures are divided into $m$ different groups, and each group is assigned a unique power level. The signal amplitudes for different numbers of active users and group sizes are listed in TABLE~\ref{tab:power ratios}, which are derived from Theorem~1 in \cite{Zhang2025VTC}.
	\begin{table}[!t]
		\centering
\caption{Signal amplitudes for different \(K_a\) and group numbers \(m\), derived from \cite[Theorem~1]{Zhang2025VTC}}
		\label{tab:power ratios}
		\renewcommand{\arraystretch}{1.1}
		\setlength{\tabcolsep}{2mm}{
			\begin{tabular}{c|c|c}
				\hline
				$K_a$ & $m$ & Power Ratios $(p_1,\ldots,p_m)$                         \\ \hline
				150               & 3                 & (0.9098,~0.9947,~1.0876)   \\
				175               & 2                 & (0.9141,~1.0791)                    \\
				200               & 2                 & (0.8997,~1.0911)                     \\
				225               & 3                 & (0.8620,~0.9878,~1.1319)                    \\
				250               & 3                 & (0.8451,~0.9847,~1.1472) \\ \hline
		\end{tabular}}
	\end{table}
	\begin{figure}[!t]
		\centering
		\includegraphics[width=\columnwidth]{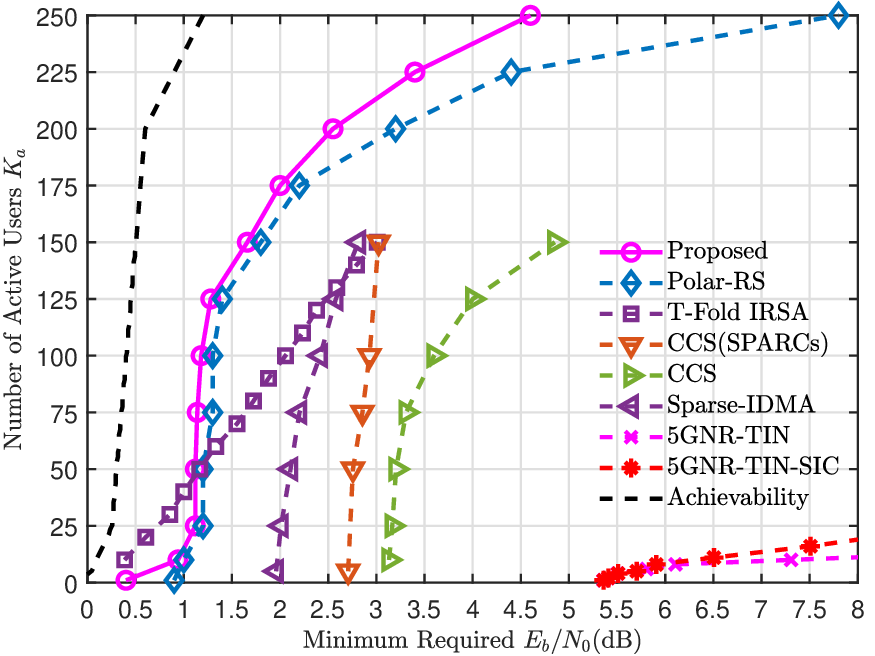}
		\caption{Performance of the minimum required energy-per-bit for a $5\times 10^{-2}$ PUPE. All benchmarks can be referred in \cite{UMA_proc}.}
		\label{fig:capacity}
	\end{figure}

	\begin{figure}[!t]
		\centering
		\includegraphics[width=\columnwidth]{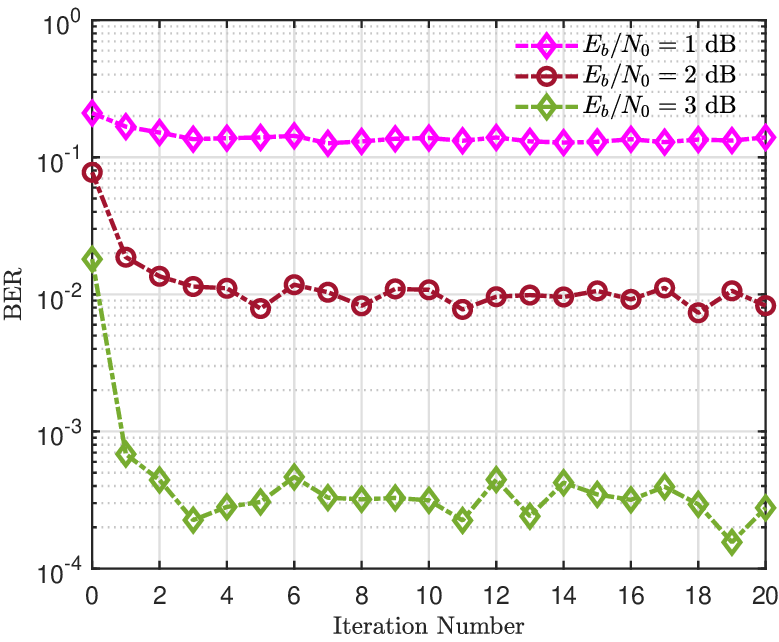}
		\caption{Illustration of convergence behavior under different energy-per-bit with $K_a=25$ and perfect signatures detection.}
		\label{fig:convergence}
	\end{figure}

 To perform a fair comparison, we select $B=97$ ($B_p=12$ and $B_c=85$) for the RS-Polar scheme in~\cite{RS-Polar} to achieve the same channel coding rate of $1/3$. The list size for the polar decoder is set to $16$. For the proposed scheme, if the output list size is smaller than the number of detected signatures after SIC, i.e., $|\tilde{\mathcal{L}}|< K_a$, the receiver then conducts MSE-based decoding in \cite{RS-Polar} for another trial.
 
 Figure~\ref{fig:capacity} illustrates the minimum energy-per-bit required to achieve a PUPE target of 0.05. The results confirm the effectiveness of the proposed iterative decoder, which outperforms many existing URA schemes. For up to 25 active users, RS schemes exhibit a sharper increase in energy requirements but stabilize quickly, showing greater resistance to changes in $K_a$ compared to T-Fold IRSA. Beyond 25 users, RS schemes maintain the most robust capacity performance as user density increases. The proposed decoder surpasses RS-Polar by 0.1–0.2 dB for up to 175 users, with the performance gap widening significantly to nearly 4 dB at 250 users, underscoring its potential. Notably, the decoder integrates seamlessly with any channel code supporting soft extrinsic and intrinsic information, and its performance can be further improved with advanced channel code designs. 
 
Furthermore, the convergence behavior directly influences the implementation overhead. Accordingly, Fig.~\ref{fig:convergence} examines the convergence behavior in terms of bit error rate (BER) with $K_a=25$ active users and perfect detection of spreading signatures. Each data point is simulated 1,000 times. Evidently, the proposed decoder achieves convergence in approximately 4 iterations, demonstrating feasibility and efficiency for practical implementation.
\section{Conclusion}\label{sec.5}
In this work, we proposed a novel RS-URA scheme that integrates random spreading with an iterative Gaussian approximation decoder and power division. The scheme effectively exploits the anti-interference benefits of spectrum spreading and achieves near-optimal performance in the low-activity regime. Simulation results confirm that the proposed decoder consistently outperforms existing URA schemes, including RS-Polar, with gains of up to 4 dB when supporting 250 active users. Moreover, the decoder architecture is compatible with any channel code that enables soft extrinsic and intrinsic information exchange, providing flexibility for future enhancements. Notably, the receiver requires only a few iterations to converge, making it well suited for practical implementation. Overall, these results demonstrate that the proposed scheme is a scalable and efficient solution for mMTC scenarios.
%This paper presented a novel RS-URA scheme that combines random spreading with an iterative Gaussian approximation decoder and power division. The proposed approach effectively harnesses the anti-interference benefits of spectrum spreading and achieves near-optimal performance in the low-activity regime. Simulation results demonstrate that the proposed decoder outperforms existing URA schemes, including RS-Polar, particularly as the number of active users increases. Specifically, it achieves up to 4~dB gain over RS-Polar at 250 users. Moreover, the decoder architecture is compatible with any channel code that supports soft extrinsic and intrinsic information exchange, offering flexibility for future enhancements. And it takes only few iterations to converge the receiver, which is favorable for practical implementation. These results highlight the potential of the proposed scheme as a scalable and efficient solution for mMTC scenarios.

\balance

\end{document}